\documentclass[12pt]{article}
\usepackage[dvips]{graphicx}
\usepackage{latexsym}
\usepackage{amssymb}
\newcommand{\NN}{\mathbb{N}}
\newcommand{\C}{\mathbb{C}}
\newcommand{\CP}{\mathbb{CP}}

\newcommand{\R}{\mathbb{R}}
\newcommand{\T}{\mathbb{T}}
\newcommand{\Z}{\mathbb{Z}}

\let\a=\alpha

\def\t{\tilde}
\def\ov{\overline}

\def\p{\partial}

\newcommand{\bea}{\begin{eqnarray}}
\newcommand{\eea}{\end{eqnarray}}

\newcommand{\koniec}{\begin{flushright}  $\Box $ \end{flushright}}
\def\be{\begin{equation}}
\def\ee{\end{equation}}

\def\theequation{\thesection.\arabic{equation}}

\def\t{\tilde}

\def\v{\bf v}
\def\u{\bf u}

\def\O{\cal O}

\def\om{\omega}

\def\ov{\overline}

\def\lt{\tilde{\lambda}}
\def\p{\partial}

\newcommand{\spp}{\mathbb{S}}
\def\ov{\overline}

\newcommand{\hook}{{\setlength{\unitlength}{11pt}   
                   \begin{picture}(.833,.8)
                   \put(.15,.08){\line(1,0){.35}}
                   \put(.5,.08){\line(0,1){.5}}
                   \end{picture}}}
\def\a{\alpha}
\def\l{\lambda}

\def\O{{\cal O}}

\newtheorem{theo}{Theorem}[section]

\begin{document}
\title{\vskip -70pt
\begin{flushright}
{\normalsize DAMTP-2009-7} \\
\end{flushright}
\vskip 80pt
{\bf Twistor Theory and Differential Equations}
\vskip 20pt}
\author{Maciej Dunajski\thanks{email m.dunajski@damtp.cam.ac.uk}\\[15pt]
{\sl Department of Applied Mathematics and Theoretical Physics} \\[5pt]
{\sl University of Cambridge} \\[5pt]
{\sl Wilberforce Road, Cambridge CB3 0WA, UK} 
}
\date{} 
\maketitle
\begin{abstract}
This is an elementary and self--contained review of twistor theory 
as a geometric tool for solving non-linear differential equations.
Solutions to soliton equations like KdV, Tzitzeica, integrable chiral
model, BPS monopole or Sine-Gordon arise from holomorphic vector bundles over
$T\CP^1$. 
A different framework is provided for the  
dispersionless analogues of soliton equations,
like dispersionless KP or $SU(\infty)$ Toda system in 2+1 dimensions.
Their solutions correspond to deformations of  (parts of) $T\CP^1$,
and ultimately to Einstein--Weyl curved geometries generalising the flat
Minkowski space. 

A number of exercises is included and
the necessary facts about vector bundles over the Riemann sphere
are summarised in the Appendix.
\end{abstract}
\newpage
\section{Introduction}
\setcounter{equation}{0}
Twistor theory was created by  Roger Penrose 
\cite{Penrose_twistor_alg} in 1967.
The original motivation was 
to unify general relativity and quantum mechanics in a non--local theory 
based on complex numbers. The application of twistor theory to 
differential equations and integrability has been an unexpected 
spin off from the twistor programme.  It has been developed over 
the last thirty years by the Oxford school of Penrose and 
Atiyah with the crucial
early input from Richard Ward \cite{Wa77,Wa85} and Nigel Hitchin \cite{H82,Hi82} and further contributions from
Lionel Mason, George Sparling, Paul Tod, Nick Woodhouse and others.

The twistor approach to  integrability is a subject of 
the monograph \cite{MW98} as well as the forthcoming book \cite{MD_book}. 
This short review is supposed to give a self--contained introduction to the 
subject. The approach will be elementary - explicit calculations
will be used in place of (often very elegant) abstract geometric 
constructions. Filling in the gaps in these calculations should be within a 
reach of a first year research student.

I thank Prim 
Plansangkate for carefully reading the manuscript and correcting several errors.
\subsection{Motivation--integral geometry}
Twistor theory is based on projective geometry and as such
has its roots in the 19th century {Klein correspondence}. 
It can also be traced back to other areas of mathematics. One such area is
a subject now known as integral geometry 
(a relationship between  twistor theory and integral geometry has  
been explored by Gindikin \cite{gindikin}). 
\subsubsection*{Radon Transform}
Integral geometry goes back to 
Radon \cite{radon} who considered the following problem:
Let  $f:\R^2\longrightarrow \R$ be a smooth function with suitable decay 
conditions at $\infty$ (for example a function of compact
support as shown below)
\begin{center}
\includegraphics[width=10cm,height=6cm,angle=0]{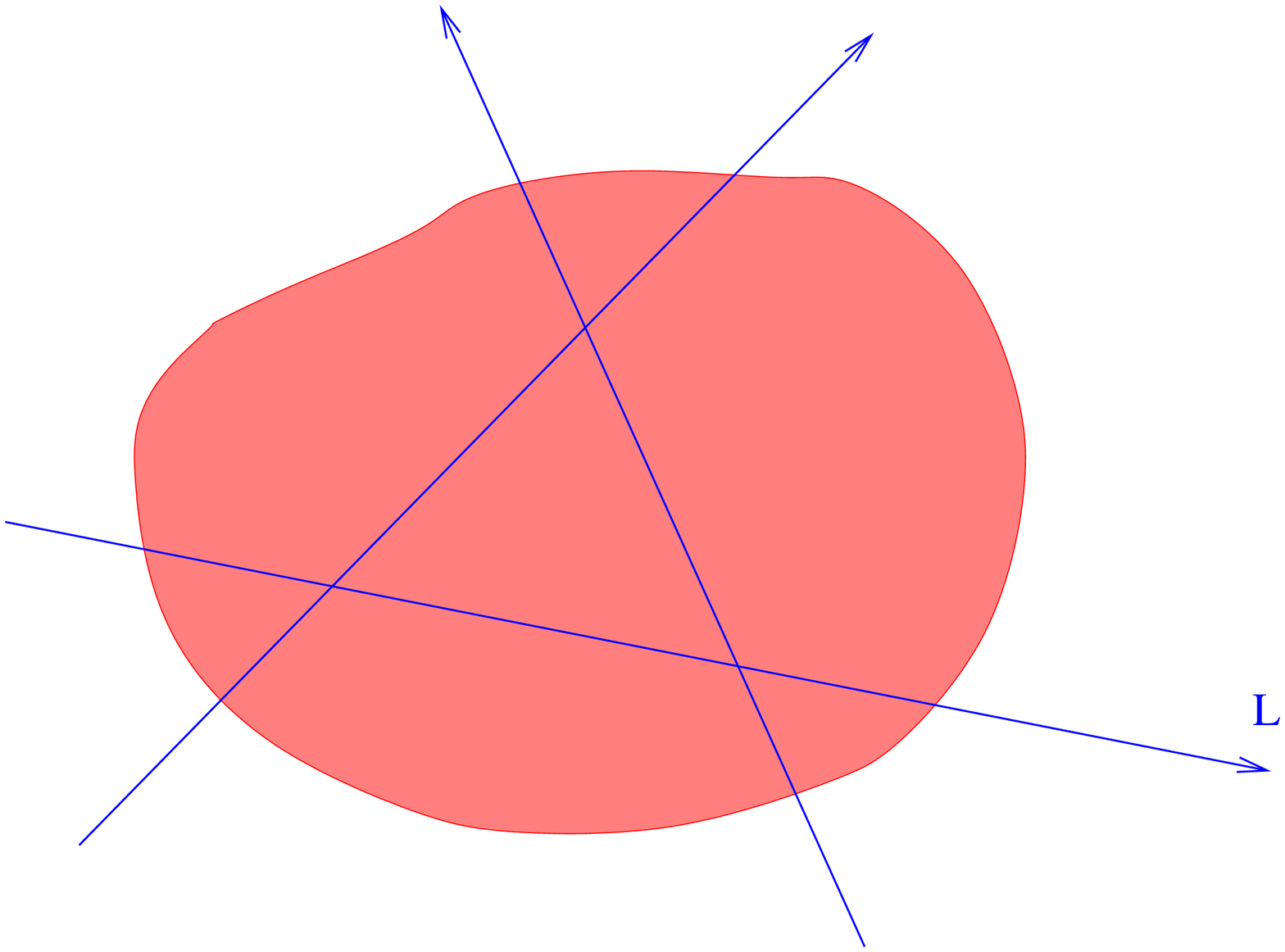}
\end{center}
and let $L\subset \R^2$ be an oriented line. Define a function
on the space of oriented lines in $\R^2$ by
\be\label{radon_1}
\phi(L):=\int_L f.
\ee
Radon has demonstrated that 
there exists an inversion formula  $\phi\longrightarrow f$.
Radon's construction can be generalised in many ways and
it will become clear  that Penrose's twistor theory is its far reaching generalisation.
Before moving on, it is however worth remarking that an extension of Radon's work
has lead to Nobel Prize awarded (in Medicine) for pure
mathematical research! It was given in 1979  to Cormack \cite{cormac}, who unaware of Radon's
results had rediscovered the inversion formula for (\ref{radon_1}),
and had explored the setup allowing the function $f$ 
to be defined on a non--simply connected region in $\R^2$
with a convex boundary. If one only allows the lines which do not pass through
the black region 
\begin{center}
\includegraphics[width=10cm,height=6cm,angle=0]{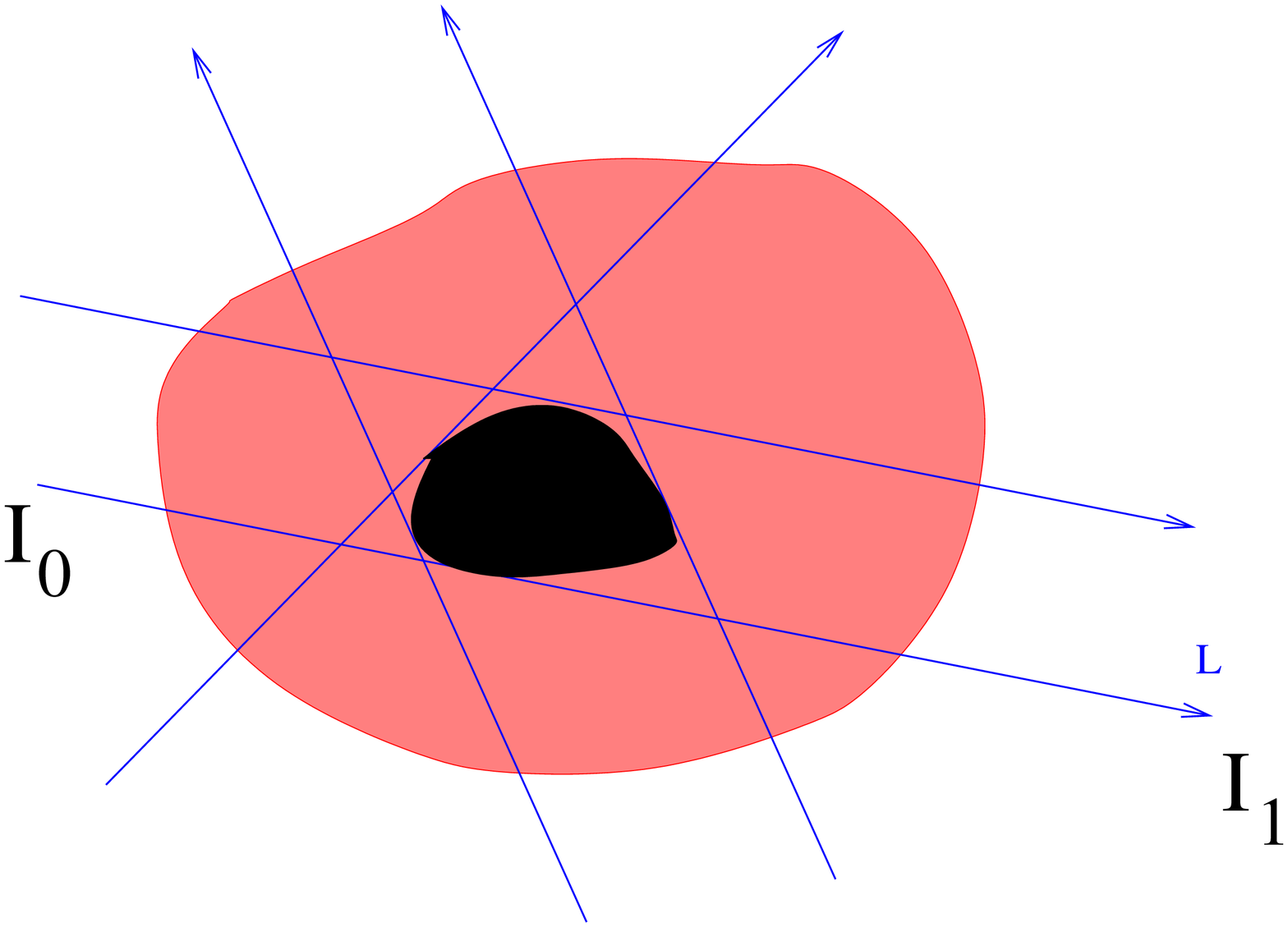}
\end{center}
or are  tangent to the boundary of this region, 
the original function $f$ may still be reconstructed from
its integrals along such lines (this is called the {\em support theorem}.
See \cite{Helgas} for details.). In the application to computer
tomography one takes a number of 2D planar section of 3D objects
and relates the function $f$ to the (unknown) density of these objects.
The input data given to a radiologist consist of the intensity of the incoming and outgoing X--rays passing through the object
with intensities $I_0$ and $I_1$ respectively
\[
\phi(L)=\int_L\frac{d I}{I}=\log{I_1}-\log{I_0} =-\int_L f
\]
where $dI/I=-f(s)ds$ is the relative infinitesimal intensity loss 
inside the body on an interval of length $ds$.

The Radon transform
then allows to recover $f$ from this data, and the generalisation
provided by the support theorem
becomes important if not all regions in the object (for example patient's
heart) can be X-rayed.
\subsubsection*{John Transform}
The inversion formula for  Radon transform (\ref{radon_1})
can exist because both $\R^2$ and the space of oriented lines
in $\R^2$ are two dimensional. Thus, at least naively, one function of two variables can be constructed from another such function 
(albeit defined on a different space). This symmetry does not hold
in higher dimensions, and this underlines the following important result
of John \cite{john}. Let $f:\R^3\longrightarrow \R$ be a function (again, 
subject
to some decay conditions which makes the integrals well defined) and let
${L}\subset \R^3$ be an oriented line.
Define $\phi({L})=\int_{L} f$, or
\be
\label{john_form}
\phi({\alpha_1, \alpha_2, \beta_1, \beta_2})=\int^{\infty}_{-\infty}
f({{\alpha_1}}s+{{\beta_1}},{{\alpha_2}} s+{{\beta_2}}, s)ds
\ee
where $(\alpha, \beta)$ parametrise the four--dimensional space
$\T$ of oriented lines in $\R^3$ (Note that this parametrisation
misses out the lines parallel to the plane $x_3=\mbox{const}$.
The whole construction can be done invariantly without choosing any 
parametrisation, but here we choose the explicit approach for clarity).
The space of oriented lines is four dimensional, and $4>3$
so expect one condition on $\phi$. Differentiating under the integral sign 
yields the  ultrahyperbolic wave equation 
\[
\frac{\partial^2\phi}{\partial \alpha_1\partial \beta_2}
-\frac{\partial^2\phi}{\partial \alpha_2\partial \beta_1}=0,
\]
and John has shown that all smooth solutions to this equation arise
from some function on $\R^3$. This is a feature of 
twistor theory: an unconstrained function  on twistor space (which in this case is identified with $\R^3$) yields
a solution to a differential equation on space--time (in this case
locally $\R^4$ with a metric of $(2, 2)$ signature).
After the change of  coordinates 
$\alpha_1=x+y, \alpha_2=t+z, \beta_1=t-z, \beta_2=x-y$ the equation becomes
\[
\frac{\partial^2\phi}{\partial x^2}
+\frac{\partial^2\phi}{\partial z^2}-\frac{\partial^2\phi}{\partial y^2}
-\frac{\partial^2\phi}{\partial t^2}=0
\]
which may be relevant to physics with two times! The integral formula
given in the next section corrects the `wrong' signature
to that of the Minkowski space and is a starting point of twistor theory.
\subsubsection*{Penrose Transform}
In 1969 Penrose gave a formula for solutions to
wave equation in Minkowski space \cite{Pen_2} 
\be
\label{penrose_form}
\phi({x,y,z,t})=\oint_{\Gamma\subset \CP^1}
f({{(z+t)}}+{{(x+i y)}}\lambda,{{(x-i y)}} 
-{{(z-t)}}\lambda, \lambda)d\lambda.
\ee
Here $\Gamma\subset \CP^1$ is a closed contour and
the function $f$ is holomorphic on $\CP^1$ except some number of poles.
Differentiating the RHS verifies  that
\[
\frac{\p^2 \phi}{\p t^2}-\frac{\p^2 \phi}{\p x^2}-
\frac{\p^2 \phi}{\p y^2}-\frac{\p^2 \phi}{\p z^2}=0.
\]
Despite the superficial similarities the Penrose formula is mathematically
much more sophisticated than John's formula (\ref{john_form}).
One  could modify a contour
and add a holomorphic function inside the contour to $f$
without changing the solution $\phi$. The proper description 
uses  {sheaf cohomology} which considers equivalence classes
of functions and contours (see e.g. \cite{WW89}).
\subsection{Twistor Programme}
Penrose's formula (\ref{penrose_form}) 
gives real  solutions to the wave equation
in Minkowski space from holomorphic functions of three arguments. According to
the twistor philosophy this appearance of complex numbers should be 
understood at a fundamental, rather than technical, level. 
In quantum physics the complex numbers are regarded as fundamental:
the complex wave function is an element of a complex Hilbert space.
In twistor theory Penrose aimed to bring the classical physics
at the equal footing, where the complex numbers play a role from the start. 
This already takes place in special relativity, where the complex numbers
appear on the celestial sphere
visible to an observer on a night sky. 
\begin{center}
\includegraphics[width=10cm,height=6cm,angle=0]{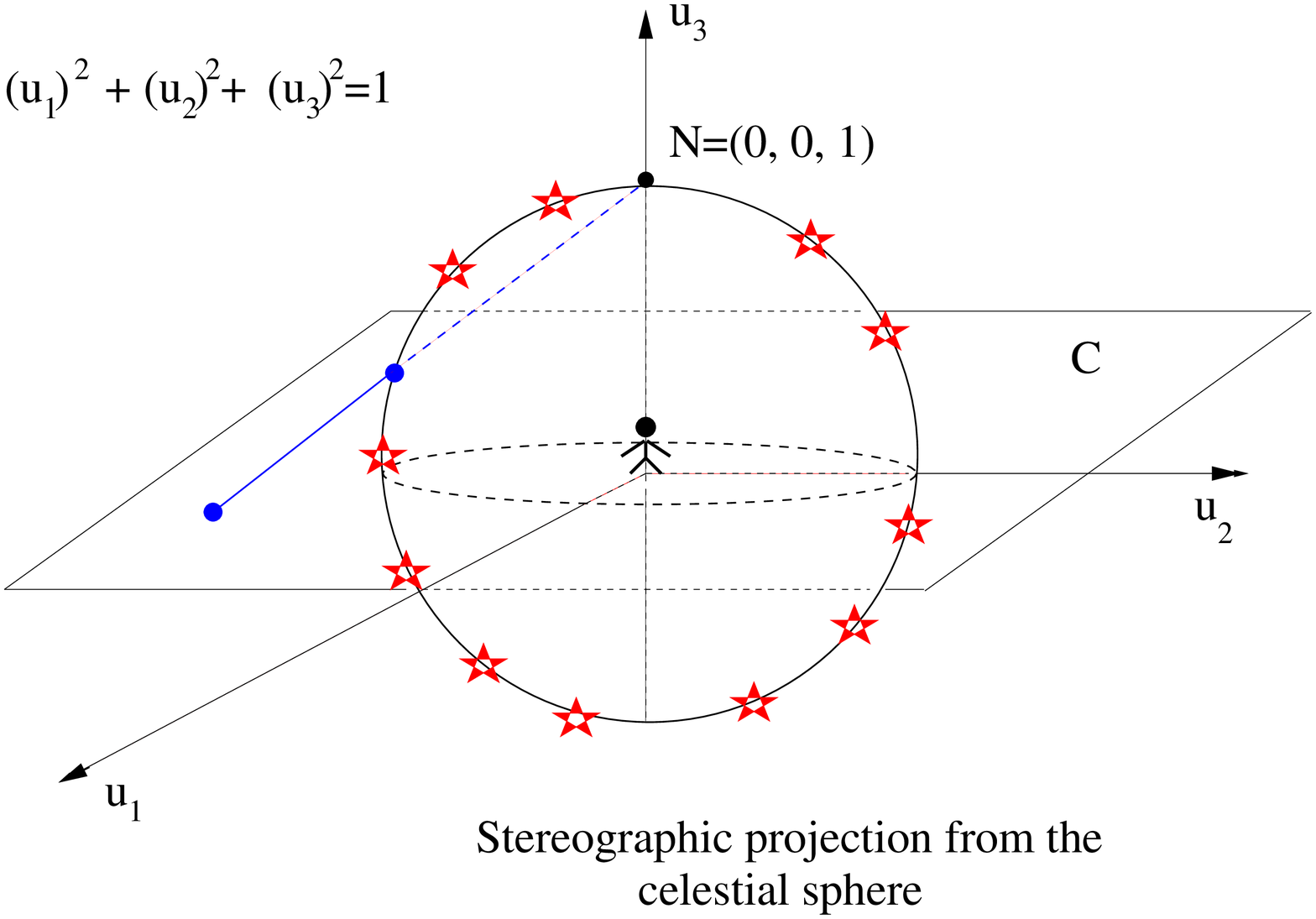}
\end{center}
The two--dimensional sphere
is the simplest example of a non--trivial complex manifold (see Appendix for more details).
Stereographic projection from the north pole $(0, 0, 1)$ gives 
a complex coordinate
\[\lambda=\frac{u_1+iu_2}{1-u_3}.
\]
Projecting
from the south pole $(0, 0, -1)$ gives another coordinate 
\[\tilde{\lambda}=\frac{u_1-iu_2}{1+u_3}.\]
On the overlap $\tilde{\lambda}=1/\lambda$. Thus the transition function is 
holomorphic and   this makes
$S^2$ into a complex manifold  $\CP^1$ (Riemann sphere).
The double covering
$SL(2, \C)\stackrel{2:1}\longrightarrow SO(3, 1)$
can be understood in this context. If worldlines of two observers
travelling with relative constant velocity intersect at
a point in space--time, the celestial spheres
these observers see are related by a M\"obius transformation
\[
\lambda\rightarrow \frac{\alpha\lambda+\beta}{\gamma\lambda+\delta}
\]
where the unit--determinant matrix
\[
\left (
\begin{array}{cc}
\alpha &\beta\\
\gamma & \delta
\end{array}
\right )
\in SL(2, \C)
\]
corresponds to the Lorentz transformation relating the two observers.

The celestial sphere is a past light cone of an observer ${\bf O}$ which consist of light rays
through an event ${\bf O}$ at a given moment. In the twistor approach the light rays are regarded as more fundamental than events in space--time. 
The five dimensional space
of light rays ${\cal PN}$  
in the Minkowski space is a hyper--surface in a three dimensional complex
manifold ${\cal PT}=\CP^3-\CP^1$ called the projective twistor 
space. ({\bf Exercise}: Why is ${\cal PN}$ five dimensional? Show that
as a real manifold ${\cal PN}\cong S^2\times \R^3$).

Let $(Z^0, Z^1, Z^2, Z^3)\sim (cZ^0, cZ^1, cZ^2, cZ^3), c\in \C^*$ 
with $(Z^2, Z^3)\neq (0, 0)$ be 
homogeneous coordinates of a twistor (a point in ${\cal PT}$). 
The twistor space and the Minkowski space are linked by
the incidence relation
\be
\label{4d_incidence}
\left (
\begin{array}{cc}
Z^0\\
Z^1 
\end{array}
\right )
=
\frac{i}{\sqrt{2}}
\left (
\begin{array}{cc}
t+z & x+iy\\
x-iy & t-z
\end{array}
\right )
\left (
\begin{array}{cc}
Z^2 \\
Z^3 
\end{array}
\right )
\ee
where $x^\mu=(t, x, y, z)$ are coordinates of a point in Minkowski space.
({\bf Exercise}: show that if two points in  Minkowski space
are incident with the same twistor, then they are null separated).
Define the Hermitian inner product 
\[
\Sigma(Z, \ov{Z})=Z^0\ov{Z^2}+Z^1\ov{Z^3}
+Z^2\ov{Z^0}+Z^3\ov{Z^1}
\]
on the non--projective twistor space ${\cal T}=\C^4-\C^2$. The signature of
$\Sigma$ is $(+ + - -)$ so that the orientation--preserving endomorphisms of 
${\cal T}$ preserving $\Sigma$ form a group $SU(2, 2)$. This group has fifteen
parameters and is locally isomorphic to the conformal group $SO(4, 2)$ of
the Minkowski space. We divide the twistor space into three parts
depending on whether $\Sigma$ is positive, negative or zero. This partition descends
to the projective twistor space. In particular
the hypersurface
\[
{\cal PN}=\{[Z]\in {\cal PT}, \Sigma(Z, \ov{Z})=0\}\subset {\cal PT}
\]
is preserved by the conformal transformations of the Minkowski space
which can be verified directly using (\ref{4d_incidence}).

Fixing the coordinates $x^\mu$ of a  space--time point 
in (\ref{4d_incidence}) gives a plane in the non--projective twistor
space $\C^4-\C^2$ or a projective line $\CP^1$ in ${\cal PT}$.
If the coordinates $x^\mu$ are real this line lies in the
hypersurface ${\cal PN}$. Conversely, fixing a twistor
in ${\cal PN}$ gives a light--ray in the Minkowski space.

So far only the null twistors (points in ${\cal PN}$) have been 
relevant in this discussion. General points in ${\cal PT}$
can be interpreted in terms of the complexified Minkowski space
$\C^4$ where they correspond to null two--dimensional planes with
self--dual tangent bi-vector. This, again, is a direct consequence of
(\ref{4d_incidence}) where now the coordinates $x^{\mu}$ are complex.
There is also an interpretation of non--null twistors in the real 
Minkowski space, but
this is far less obvious \cite{Penrose_twistor_alg}: The Hermitian 
inner product
$\Sigma$ defines a vector space ${\cal T}^*$ dual to the 
non--projective twistor space.
The elements of the corresponding projective space ${\cal PT}^*$ are called dual twistors.
Now take a non--null twistor $Z\in {\cal PT}$. Its dual $\ov{Z}\in{\cal PT}^*$ corresponds to a projective
two plane $\CP^2$ in ${\cal PT}$. ({\bf Exercise}: Use (\ref{4d_incidence}) to find
an explicit equation for this plane). A holomorphic two--plane intersects the hyper--surface
${\cal PN}$ in a real three--dimensional locus. This locus corresponds to
a three--parameter family of light--rays in the real Minkowski space. 
This family
representing a single twistor is called the Robinson congruence. 
A picture of this configuration which appears on the front cover of \cite{PR86}
shows a system of twisted oriented circles in the Euclidean space $\R^3$, the point 
being that
any light--ray is represented by a point in $\R^3$ together with an arrow 
indicating
the direction of the ray's motion. This configuration 
originally gave
rise to a name `twistor'. 

 Finally we can give a twistor interpretation
of the contour integral formula (\ref{penrose_form}). Consider a function
$f=f(Z^0/Z^2, Z^1/Z^2, Z^3/Z^2)$  which is holomorphic on an intersection 
of two open sets covering $\cal PT$ (one of this sets is defined 
by $Z^2\neq 0$ and the other by $Z^3\neq 0$) and restrict this function to a 
rational curve (\ref{4d_incidence}) in ${\cal PN}$. Now integrate $f$ along
a contour in this curve. This gives (\ref{penrose_form}) with 
$\lambda =Z^3/Z^2$. ({\bf Exercise}: Explain why $f$, when viewed as a function
on the non--projective twistor space, must be homogeneous of degree
$-2$ in $Z^{\alpha}$. Find a solution $\phi$ to the wave equation 
corresponding to $f=(A_\alpha Z^{\alpha})^{-1}(B_\beta Z^{\beta})^{-1}$,
where $\alpha, \beta= 0, \dots, 3$ and $(A_\alpha, B_\beta)$ are
constant complex numbers).

\vskip5pt
To sum up, the space-time points are derived objects in twistor theory.
They become `fuzzy' after quantisation. This may provide an 
attractive framework for quantum 
gravity, but it must be said that despite  40 years of research 
the twistor theory is still waiting  to have its major impact on physics.
It has however had surprisingly major impact on pure mathematics: 
ranging from representation theory and  
differential geometry to  solitons, instantons and integrable systems.

This ends the `historical' part of the review. The rest of the review
is intended to give a `down-to-earth' introduction to the calculations
done in twistor theory. Rather than using the twistors of 3+1 dimensional 
Minkowski space, we shall focus on mini--twistors which arise
in 2+1 dimensional Minkowski space or in $\R^3$.
This `mini--twistor theory'
 is in many ways simpler but still 
sufficient in applications to 2+1 and 3 dimensional integrable systems and their reductions.
The mini--twistor space $\T$
(from now on just called the twistor space) is the holomorphic tangent bundle to the Riemann
sphere. The difference between the Lorentzian and Euclidean signature of the corresponding
space--time is encoded in the anti--holomorphic involution on $\T$ which,
when restricted to rational curves, becomes the antipodal map in the Euclidean case
and the equator-fixing conjugation in the Lorentzian case. We shall study the Euclidean theory
in the next Section and the Lorentzian theory in Section \ref{sec_lor}.

\section{Non--abelian monopoles and Euclidean mini--twistors}
It is well known that the problem of finding harmonic functions in $\R^2$ 
can be solved `in one line' by introducing complex numbers: Any solution
of a two--dimensional Laplace equation $\phi_{xx}+\phi_{yy}=0$ 
is a real part of a function holomorphic in $x+iy$. This technique
fails when applied to the Laplace equation in three dimensions
as $\R^3$ can not be identified with $\C^n$ for any $n$.

Following Hitchin \cite{H82} we shall  associate a two--dimensional 
complex manifold to the three--dimensional Euclidean  space.
Define the twistor space $\T$ to be the space of oriented lines in $\R^3$.
Any oriented line is of the form ${\v}+s{\u}, \quad s\in \R$
where $\u$ is a unit vector giving the direction of the line, and $\v$
is orthogonal to $\u$ and joints the line with some chosen point
(say the origin) in $\R^3$.
\begin{center}
\includegraphics[width=10cm,height=6cm,angle=0]{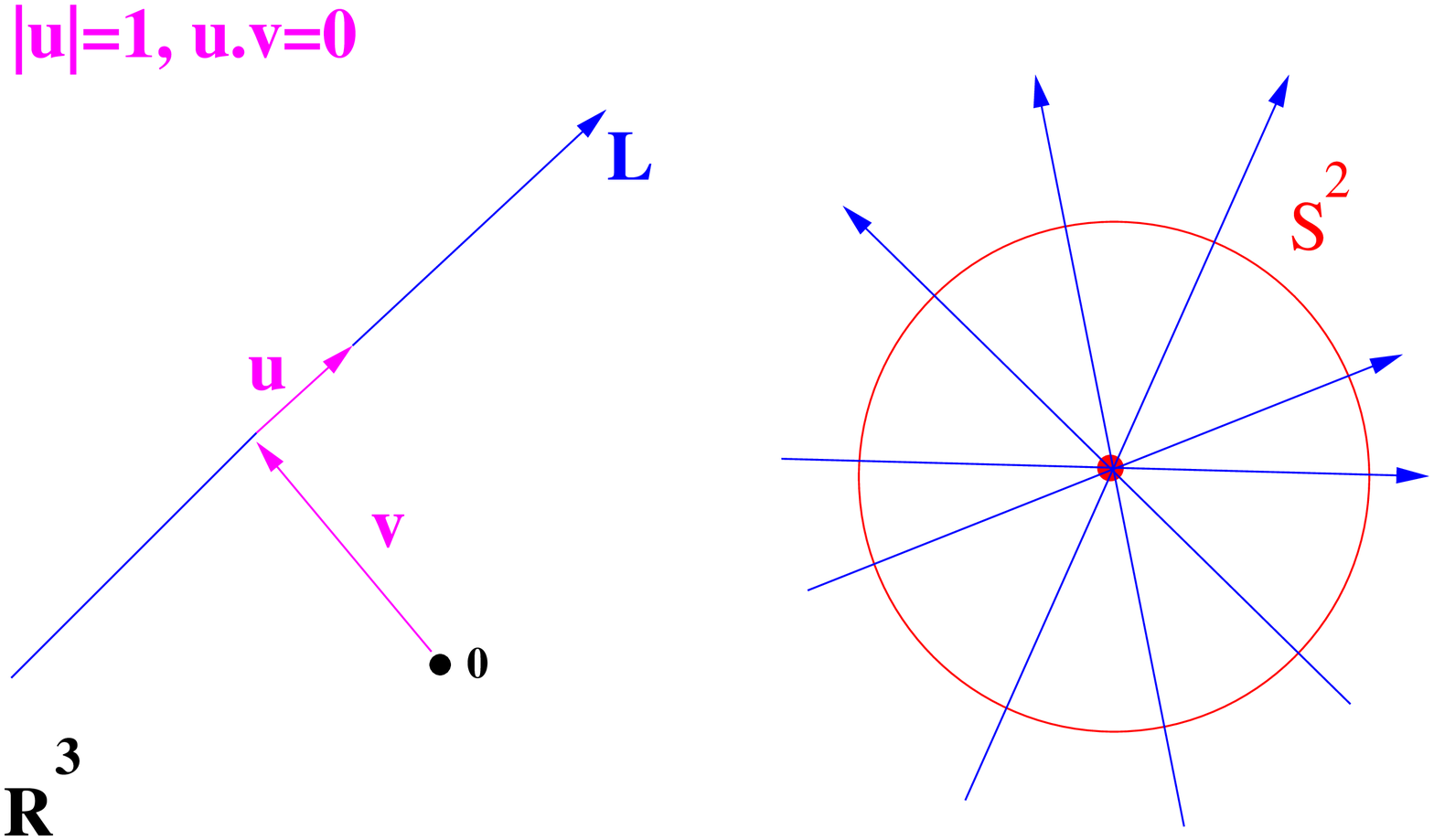}
\end{center}
Thus
\[
{\T}=\{({\u}, {\v})\in S^{2}\times\R^{3}, \;{\u.\v} =0 \}
\]
and the dimension of $\T$ is four. For each fixed ${\u}\in S^2$ this space restricts to a tangent plane to $S^2$.  The twistor space is the union of 
all tangent planes -- the tangent bundle $TS^{2}$.
This is a  topologically nontrivial manifold: Locally it is diffeomorphic to 
$S^2\times\R^2$ but globally it is twisted
in a way analogous to M\"obius strip. 
\begin{center}
\includegraphics[width=6cm,height=3cm,angle=0]{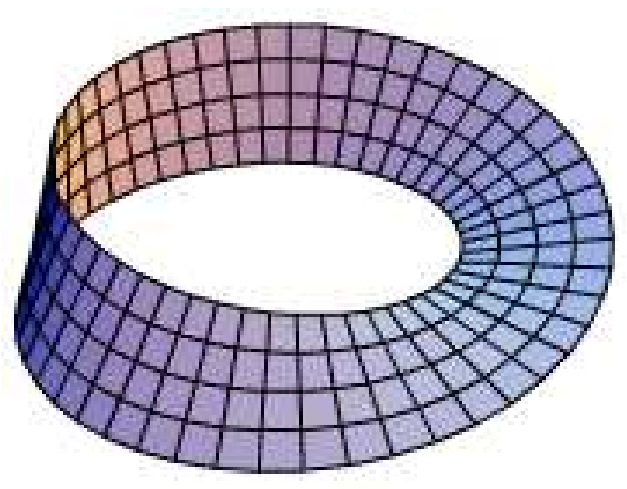}
\end{center}
Reversing the orientation of lines induces a map 
$\tau:{\T}\longrightarrow{\T}$ given by
\[
\tau(\u, \v)=(-\u, \v).
\]
The points ${\bf p}=(x, y, z)$ in $\R^3$ correspond to two--spheres in $\T$
given by  $\tau$--invariant maps
\be
\label{real_points}
\u\longrightarrow ({\u, {\bf v(u)}={\bf p}}-{({\bf p}.\u)\u})
\in\T
\ee
which are sections of the projection $\T\rightarrow S^2$.
\subsubsection*{Twistor space as a complex manifold}
Introduce the local holomorphic coordinates on an open set $U\subset \T$ 
where ${\u}\neq(0, 0, 1)$ by 
\[
\lambda=\frac{u_1+iu_2}{1-u_3} \in\CP^1=S^2,\quad \eta=\frac{v_1+iv_2}{1-u_3}+
\frac{u_1+iu_2}{{(1-u_3)}^2}v_3,
\]
and analogous complex coordinates 
$(\tilde{\lambda}, \tilde{\eta})$ in an open set $\tilde{U}$ containing ${\u}=(0, 0, 1)$. 
On the overlap 
\[
\tilde{\lambda}=1/\lambda,\qquad \tilde{\eta}=-\eta/\lambda^2.
\]
({\bf Exercise}: Work out the details of this). This endows $\T$ with a structure of complex manifold $T\CP^1$. It is a holomorphic tangent bundle to the Riemann sphere (see Appendix).
\begin{center}
\includegraphics[width=15cm,height=6cm,angle=0]{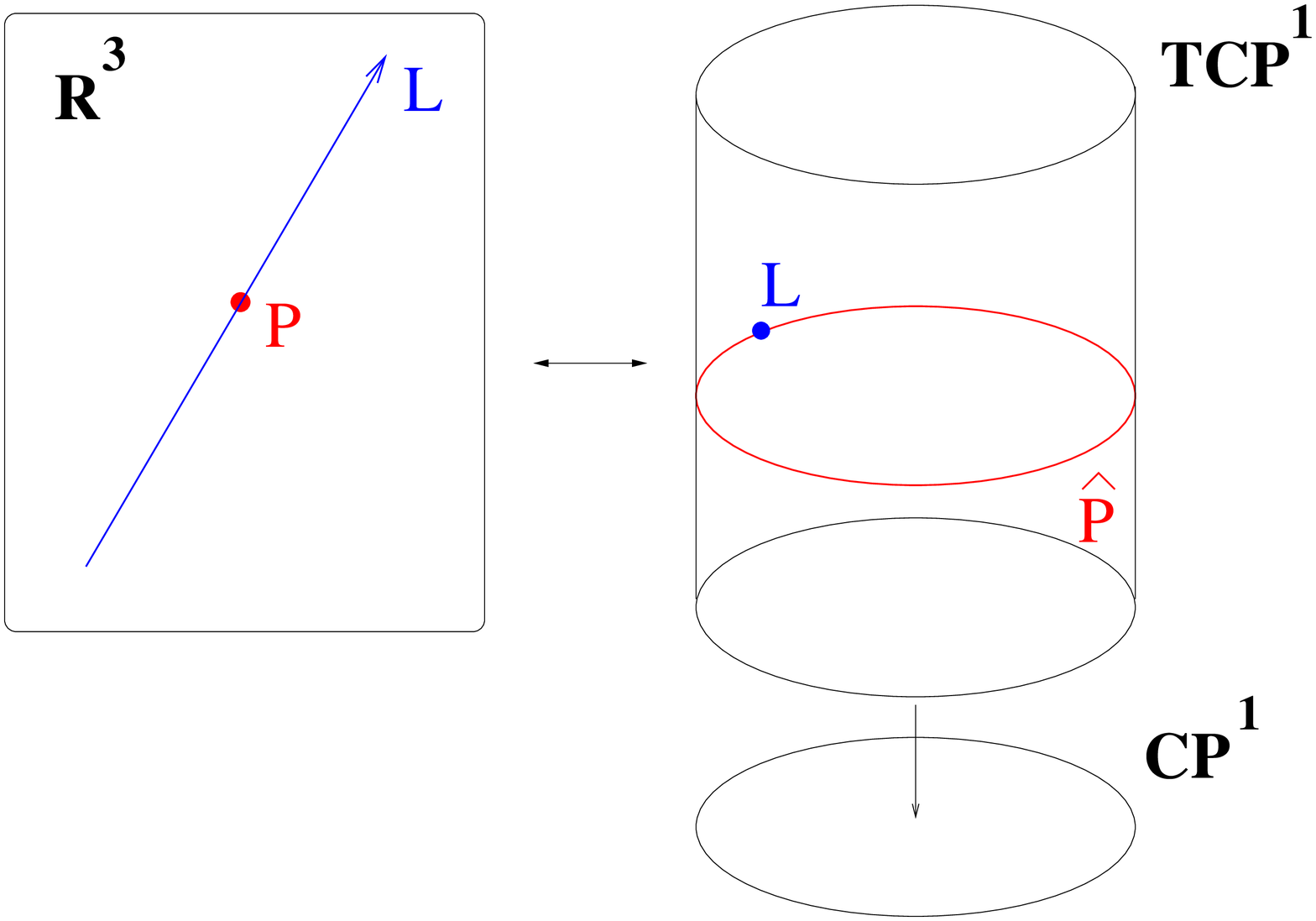}
\end{center}
In the holomorphic coordinates the line orientation reversing involution $\tau$ is given by
\be
\label{eucl_invo}
\tau(\lambda, \eta)=\Big(-\frac{1}{\ov{\lambda}},  
-\frac{\ov{\eta}}{\ov{\lambda}^2}\Big).
\ee
This is an antipodal map lifted from a two--sphere to the total space
of the tangent bundle.
The formula (\ref{real_points}) implies that
the points in $\R^3$ are $\tau$-invariant 
holomorphic maps $\CP^1\rightarrow T\CP^1$  given by
\be
\label{hol_sec_point}
\lambda\rightarrow  (\lambda, \eta={(x+iy)}+2\lambda 
{z}- \lambda^2{(x-iy)}).
\ee
({\bf Exercise}: Verify that (\ref{hol_sec_point}) follows from 
(\ref{real_points})).
\subsubsection*{Harmonic functions and abelian monopoles}
Finally we can return to our original problem.
To find a harmonic function at ${P=(x, y, z)}$
\begin{enumerate}
\item
Restrict a twistor function $f(\lambda, \eta)$ 
defined on $U\cap \tilde{U}$
to 
a line (\ref{hol_sec_point})
${\hat{P}=\CP^1=S^2}$. 
\item
Integrate along a closed contour
\be
\label{whitt_form}
{\phi(x, y, z)}=
\oint_{\Gamma\subset{\hat{P}}} 
f(\lambda, {(x+iy)}+2\lambda 
{z}-
\lambda^2{(x-iy)})d \lambda,
\ee
\item Differentiate under the integral to verify
\[
\frac{\p^2 \phi}{\p x^2}+\frac{\p^2 \phi}{\p y^2}
+\frac{\p^2 \phi}{\p z^2}=0.
\]
This formula was already known to Whittaker \cite{whittaker} in 1903,
albeit Whittaker's formulation does not make any use of complex numbers 
and his formula is given in terms of a real integral.
\end{enumerate}
Small modification of this formula can be used to solve a 1st order linear
equation for a function $\phi$ and a magnetic potential 
${\bf A}=(A_1, A_2, A_3)$ of the form
\[
\nabla\phi=\nabla\wedge{\bf A}.
\]
This is the abelian monopole equation. Geometrically, the one--form
$A=A_jdx^j$ is a connection on a $U(1)$ principal bundle over $\R^3$,
and $\phi$ is a section of the adjoint bundle.
Taking the ${\bf curl}$ of both sides of this equation
implies that $\phi$ is harmonic, and conversely given a harmonic function
$\phi$ locally one can always find a one--form $A$ (defined up to 
addition of a gradient of some function) 
such that the abelian monopole equation 
holds. ({\bf Exercise}: Find an integral formula
for the one--form $A$ analogous to (\ref{whitt_form}). This question
is best handled using the spinor formalism introduced in Section 
\ref{null_subs}).

\subsection{Non--abelian monopoles and Hitchin correspondence}
Replacing $U(1)$ by a non--abelian Lie group 
generalises this picture to some equations on $\R^3$ in the following way:
Let
$(A_j, \phi)$  be anti--hermitian traceless $n$ by $n$ matrices
on  $\R^3$.  Define the  non--abelian  magnetic field 
\[
F_{jk}=\frac{\p A_k}{\p x^j}-\frac{\p A_j}{\p x^k}+[A_j, A_k], \quad
j,k=1, 2, 3.
\]
The non--abelian monopole equation is a system of non--linear PDEs
\be
\label{monopole_R3}
\frac{\p \phi}{\p x^j}+[A_j, \phi]=\frac{1}{2}\varepsilon_{jkl}F_{kl}.
\ee
These are three equations for three unknowns as
$(A, \phi)$ are defined up 
to gauge transformations
\be
\label{gauge_trans_Ward_m}
A\longrightarrow gAg^{-1} -d g \; g^{-1}, \qquad
\phi\longrightarrow g\phi g^{-1}, \qquad g=g(x, y, t)\in SU(n)
\ee
and one component of $A$ (say $A_1$) can always be set to zero.

The twistor solution to the monopole equation
consists of the following steps \cite{H82}
\begin{itemize}
\item Given  $(A_j({\bf x}), \phi({\bf x}))$ solve a matrix ODE along each
oriented line ${\bf x}(s)={\bf v}+{s\bf u}$
\[
\frac{d V}{d s}+(u^jA_j+i\phi)V=0.
\]
Space of solutions at $p\in\R^3$ is a complex vector space $\C^n$.
\item This assigns a complex vector space $\C^n$ to each point of $\T$, 
thus giving rise to a 
complex vector bundle over $\T$
with patching matrix $F(\lambda, \overline{\lambda}, \eta, \overline{\eta}) \in GL(n, \C)$.
\vskip10pt
\begin{center}
\includegraphics[width=10cm,height=6cm,angle=0]{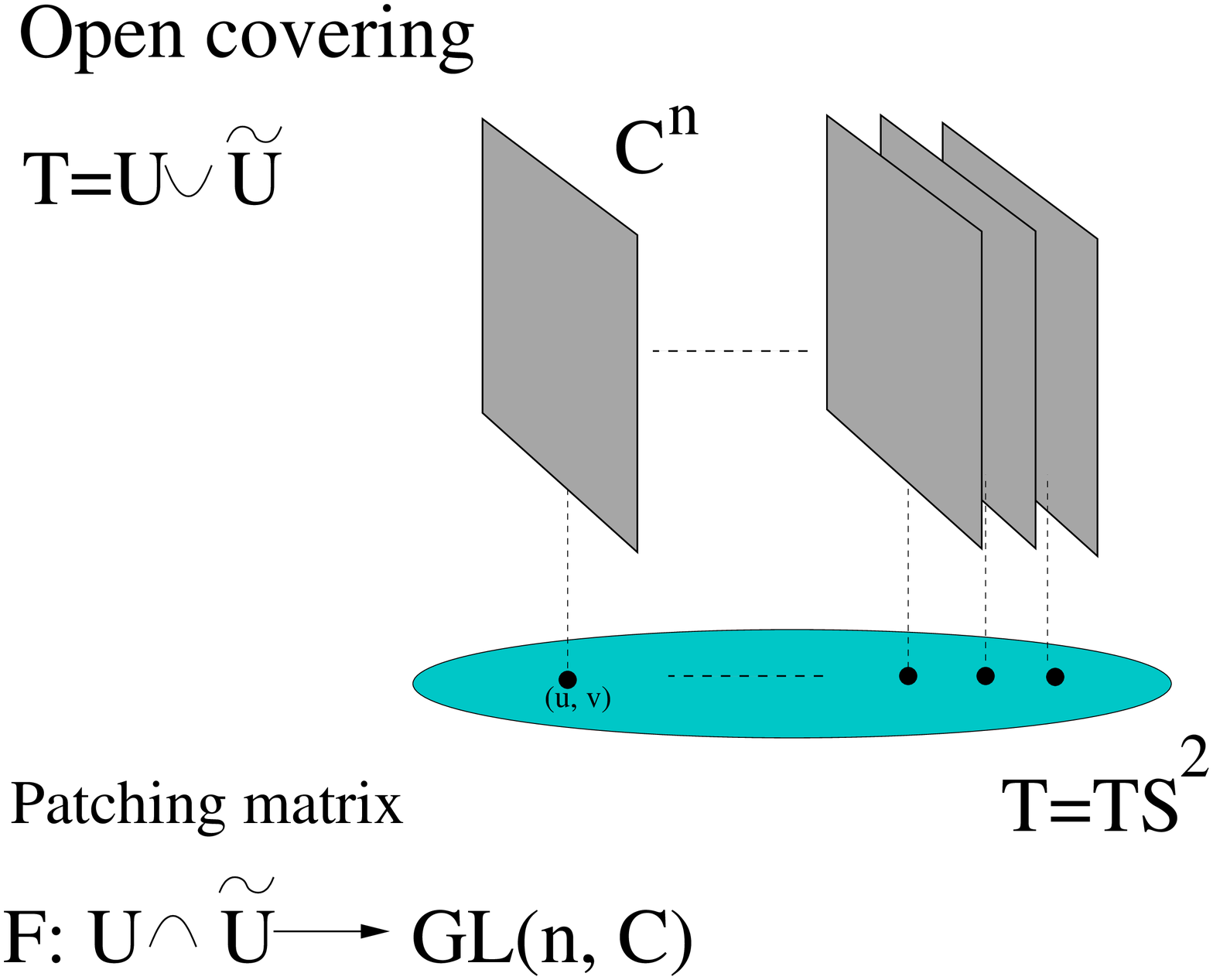}
\end{center}
\item
The
monopole equation (\ref{monopole_R3}) on $\R^3$ holds if and only if
this vector bundle is holomorphic, i.e. the 
Cauchy--Riemann equations 
\[\frac{\p F}{\p\overline{\lambda}}=0,\qquad \frac{\p F}{\p\overline{\eta}}=0
\]
hold. 
\item Holomorphic vector bundles over $T\CP^1$ are  well understood. 
Take one and work backwards to construct a monopole.
We shall work through the details of this reconstruction (albeit in complexified settings)
in the proof of Theorem \ref{main_twistor_theorem}.
\end{itemize}
\section{The Ward model and Lorentzian mini--twistors}
\label{sec_lor}
In this Section we shall demonstrate how mini--twistor
theory can be used to solve non--linear equations in $2+1$ dimensions.
Let $A=A_\mu dx^\mu$ and $\phi$ be a one--form 
and a function respectively on the 
Minkowski space $\R^{2, 1}$ with values in a Lie algebra
of the general linear group. They are defined up 
to gauge transformations (\ref{gauge_trans_Ward_m})
where $g$ takes values in $GL(n, \R)$.

Let $D_\mu=\p_\mu+A_\mu$ be a covariant derivative, and
define  $D\phi=d\phi+[A, \phi]$. 
The Ward model is a system of PDEs (\ref{monopole_R3})
where now the indices are raised  using the metric  on $\R^{2, 1}$.
If the metric and the volume form
are chosen to be
\[
h=dx^2 -4dudv , \qquad \mbox{vol}=du\wedge dx\wedge dv
\]
where the coordinates $(x, u, v)$ are real
the equations become
\be
\label{Wardeq0}
D_x\phi=\frac{1}{2}F_{uv}, \qquad D_u \phi=F_{ux}, \qquad D_v \phi=F_{xv}
\ee
where $F_{\mu\nu} = [D_\mu, D_\nu]$. These  equations arise as the
integrability conditions for an overdetermined system of linear
Lax equations
\be
\label{laxpairW}
L_0\Psi=0, \quad L_1\Psi=0, \quad\mbox{where} \quad
L_0=D_u-\l(D_x+\phi),\, L_1=D_x-\phi-\l D_v,
\ee
and $\Psi=\Psi(x, u, v, \lambda)$ takes values in $GL(n , \C)$.
We shall follow \cite{W89} and `solve' the system by establishing one--to--one correspondence between its solutions and
certain holomorphic vector bundles over the twistor space
$\T$. This construction is of interest in soliton theory as
many known integrable models
arise as symmetry reduction and/or choosing a gauge in 
(\ref{Wardeq0}).
To this end we note a few examples of such reductions. 
See \cite{MW98}
for a much more complete list.
\begin{itemize}
\item
Choose the unitary gauge group $G=U(n)$.
The integrability conditions for (\ref{laxpairW}) imply the
existence of a gauge $A_v=0$, and  $A_x=-\phi$, and a matrix
$J:\R^{2,1}\longrightarrow U(n)$ such that
\[
A_u=J^{-1} \p_u J, \qquad
A_x=-\phi=\frac{1}{2}J^{-1}\p_x J.
\]
With this gauge choice the equations (\ref{Wardeq0}) become
the integrable chiral model 
\be \label{Wardeq}
\p_v(J^{-1}\p_u J)- \p_x(J^{-1}\p_x J)=0.
\ee
This formulation breaks the Lorentz invariance of (\ref{Wardeq0})
but it allows an introduction of a positive definite energy functional.
See \cite{W88} where more details can be found.
\item Solutions to  equation
(\ref{Wardeq0})  with the gauge group  $SL(2, \R)$ which are
invariant
under a  null translation given by a Killing vector 
$K$ such that the matrix $K\hook A$
is nilpotent are characterised by the KdV equation
\cite{MS89}.
\item 
The direct calculation shows that the Ward equations with the gauge group $SL(3, \R)$
are solved by
the ansatz
\begin{eqnarray}
\label{Tansatz}
\phi&=&\frac{1}{2}\left (
\begin{array}{ccc}
0&0&1\\
0&0&0\\
-e^{\psi}&0&0
\end{array}
\right ),\\
A&=&\frac{1}{2}\left (
\begin{array}{ccc}
0&0&1\\
0&0&0\\
e^{\psi}&0&0
\end{array}
\right )dx+
\left (
\begin{array}{ccc}
\psi_u&0&0\\
 1&-\psi_u&0\\
 0&1&0
\end{array}
\right ) du+
\left (
\begin{array}{ccc}
0&e^{-2\psi}&0\\
0&0&e^{\psi}\\
0&0&0
\end{array}
\right )dv\nonumber
\end{eqnarray}
iff $\psi(u, v)$ satisfies the  Tzitz\'eica equation 
\be
\label{tzitzeica}
\frac{\p^2\psi}{\p u\p v}=e^\psi-e^{-2\psi}.
\ee
This reduction can also be characterised in a gauge invariant
manner using the Jordan normal forms for the Higgs fields. See
\cite{DP08} for details.
({\bf Exercise}: Show that  (\ref{tzitzeica}) follows from 
(\ref{Wardeq0}). What can you say about the gauge field corresponding to 
the trivial solution $\psi=0$?).
\end{itemize}
\subsection{Null planes and Ward correspondence}
\label{null_subs}
The geometric interpretation of the Lax representation
(\ref{laxpairW}) is the following. For any fixed pair of real numbers
$(\eta, \lambda)$ the plane
\be
\label{twistor_curve}
\eta=v+x\lambda+ u\lambda^2
\ee
is null with respect to the Minkowski metric on $\R^{2, 1}$, and conversely 
all null planes can be put in this form if one allows $\lambda=\infty$. 
The two vector
fields 
\be
\label{mini_dist}
\delta_0=\p_u-\lambda\p_x, \quad \delta_1=\p_x-\lambda \p_v
\ee
 span
this null plane. Thus the Lax equations (\ref{laxpairW}) imply that
the generalised connection $(A, \phi)$ is flat on null planes.
This underlies the twistor approach \cite{W89}, where one works
in a complexified Minkowski space $M=\C^3$, and interprets
$(\eta, \lambda)$ as coordinates in a patch of 
the twistor space
${\T}=T\CP^1$, with $\eta\in \C$ being a coordinate on the fibers
and $\lambda\in \CP^1$ being an affine coordinate on the base.
We shall adopt this complexified point of view from now on.

It is convenient to make use of the spinor formalism
based on the isomorphism
\[
TM=\spp\odot\spp
\]
where $\spp$ is rank two complex vector bundle (spin bundle) over
$M$ and $\odot$ is the symmetrised tensor product. 
The fibre coordinates of this bundle are
denoted by $(\pi^0, \pi^1)$ and the sections $M\rightarrow \spp$ are called
spinors. 
We shall regard
$\spp$ as a symplectic bundle  
with anti-symmetric product
\[
\kappa \cdot\rho=\kappa^0\rho^1-\kappa^1\rho^0=\varepsilon(\kappa, \rho)
\]
on its sections. 
The constant symplectic form $\varepsilon$ is represented by a matrix
\[
\varepsilon_{AB}=\left (
\begin{array}{cc}
0&1\\
-1&0 
\end{array}
\right).
\]
This gives an isomorphism between $\spp$ and its dual bundle, 
and thus can be used to `rise and lower the indices' according to 
$\kappa_{A}=\kappa^B\varepsilon_{BA}, \kappa^A=\varepsilon^{AB}\kappa_B$,
where $\varepsilon_{AB}\varepsilon^{CB}$ is an identity endomorphism.

Rearrange the space time coordinates $(u, x, v)$ of a displacement vector 
as a symmetric 
two-spinor
\[
x^{AB}:=
\left (
\begin{array}{cc}
u&x/2\\
x/2&v
\end{array}
\right ),
\]
such that the space-time metric is
\[
h=-2 d x_{AB} d x^{AB}.
\]
The twistor space of $M$ is the two-dimensional complex manifold 
${\T}=T\CP^1$.
Points of ${\T}$ correspond to null 2-planes in $M$ via the 
incidence relation
\be
\label{minit1}
x^{A B}\pi_{A }\pi_{B}=\om.
\ee
Here $(\om, \pi_{0}, \pi_{1})$ are homogeneous coordinates on 
${\T}$ as  $(\om, \pi_{A })\sim (c^2 \om, c\pi_{A})$, 
where $c \in \C^*$.
In the affine coordinates  $\l:=\pi_{0}/\pi_{1}, 
\eta:=\om/(\pi_{1})^2$
equation (\ref{minit1}) gives (\ref{twistor_curve}).

The projective spin space $P(\spp)$
is the complex projective line $\CP^1$.
The homogeneous coordinates are denoted by $\pi_{A}=(\pi_{0}, \pi_{1})$,
and the  two set covering of $\CP^1$ lifts to a covering of the twistor space
$\T$
\be
\label{covering}
U=\{(\omega, \pi_{A}), \pi_{1}\neq 0\}, \qquad \t {U}=\{(\omega, \pi_{A}), \pi_{0}\neq 0\}.
\ee
The functions $\l=\pi_{0}/\pi_{1}, 
\lt=1/\l$ are  the inhomogeneous coordinates in $U$ and $\t U$ respectively. 
It then follows that $\l=-\pi^{1}/\pi^{0}$.

Fixing $(\om , \pi_{A})$ gives a null plane in $M$. An alternative 
interpretation of (\ref{minit1}) is to fix $x^{AB}$.
This determines $\om$ as a function of $\pi_{A}$ i.e. a section of
$\T\rightarrow \CP^1$ when factored out by the relation
$(\om, \pi_{A})\sim(c^2 \om, c\pi_{A})$. 
These are embedded rational curves with self--intersection number $2$,
as infinitesimally perturbed curve $\eta+\delta\eta$ with $\delta\eta
= \delta v+\lambda\delta x + \lambda^2\delta u $  generically intersects 
(\ref{twistor_curve}) at two points. Two curves intersect
at one point if the corresponding points in $M$ are null separated.
This defines a conformal structure on $M$.
\begin{center}
\includegraphics[width=10cm,height=6cm,angle=0]{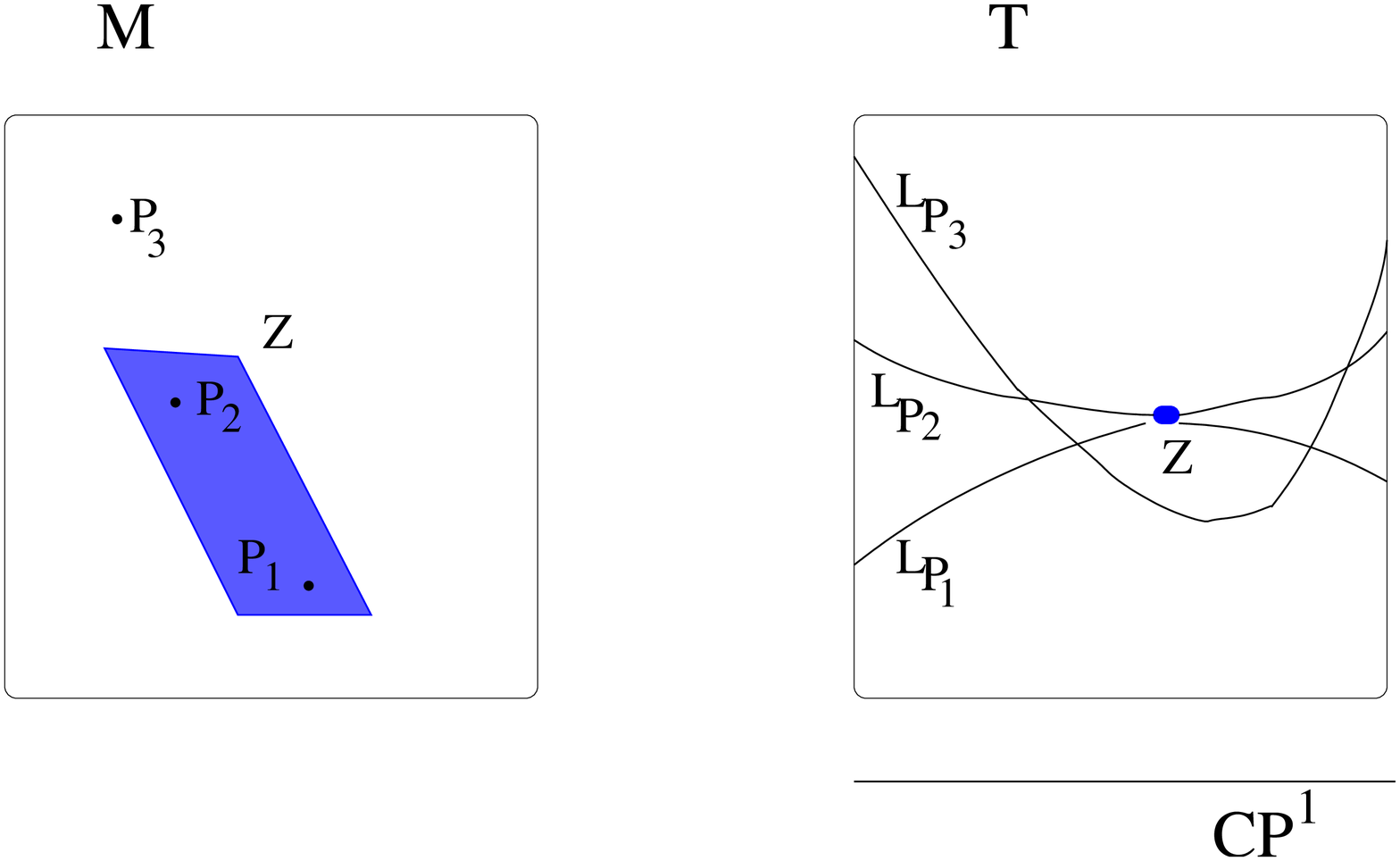}
\end{center}

The space of holomorphic sections of $\T\rightarrow \CP^1$ is $M=\C^3$
(see Appendix). The real
space-time $\R^{2+1}$ arises as the moduli space of those sections  
that are invariant under the conjugation 
\be
\label{lor_invo}
\tau(\omega,\pi_{A})=
(\bar{\omega},\bar{\pi}_{A}),
\ee
which corresponds to real $x^{AB}$. The points in $\T$ fixed by $\tau$ correspond to real
null planes in $\R^{2, 1}$. ({\bf Exercise:} Show that as a complex manifold
$\T$ is biholomorphic with a cone in $\CP^3$ with its vertex removed,
where the points in $M$ correspond to the conic sections omitting the vertex.
Demonstrate that allowing the conic sections passing through
the vertex of the cone results in a compactification
of the complexified Minkowski space $\overline{M}=M+\CP^2=\CP^3$).

\vskip5pt
The following result makes the mini--twistors worthwhile
\begin{theo}[Ward \cite{W89}]
\label{main_twistor_theorem}
There is a one--to--one correspondence between:
\begin{enumerate}
\item The gauge equivalence classes
of complex solutions to (\ref{Wardeq0}) in complexified Minkowski space
$M$
with the gauge group $GL(n, \C)$.
\item Holomorphic rank $n$ vector bundles ${E}$ over the twistor space
${\T}$ which are trivial 
on the holomorphic 
sections of $T\CP^1\rightarrow \CP^1$
\end{enumerate}
\end{theo}
{\bf Proof.} Let $(A, \phi)$ be a solution to (\ref{Wardeq0}) . 
Therefore we can integrate a pair of linear PDEs $L_0 V=L_1V=0$, 
where $L_0, L_1$ are given by (\ref{laxpairW}).
This assigns an $n$-dimensional vector space to each null plane
$Z$ in complexified Minkowski space , and so to each point $Z\in{\T}$. 
It is a fibre of a holomorphic 
vector bundle $\mu:{E}\rightarrow \T$. The bundle ${E}$ is trivial on
each section, since we can identify fibres of ${E}|_{L_p}$ at $Z_1, Z_2$
because  covariantly constant vector 
fields at null planes $Z_1, Z_2$ coincide at a common point $p\in M$.

Conversely, assume that we are given a holomorphic vector bundle  $E$ over 
${\T}$ which is trivial on each section.
Since $E|_{L_{p}}$ is trivial, and $L_p\cong \CP^1$, 
the Birkhoff-Grothendieck theorem (Appendix) gives
\[
\label{triv_on_sec}
E|_{L_p}=\O\oplus\O\oplus \dots\oplus \O
\]
and the space of sections of $E$ restricted to $L_p$ is 
$\C^n$.
This gives us a holomorphic 
rank $n$ vector bundle 
$\hat{E}$ over the complexified three--dimensional Minkowski space. 
We shall give a concrete method of
constructing a pair $(A, \phi)$ on this bundle which satisfies (\ref{Wardeq0}).

Let us cover the twistor space with two open sets $U$ and $\tilde{U}$ 
as in (\ref{covering}).
Let \[
\chi:\mu^{-1}(U)\rightarrow U\times\C^n, 
\qquad \t{\chi}:\mu^{-1}(\t U)\rightarrow {\t U}\times\C^n
\]
be local trivialisations of ${E}$, and let
$F=\t\chi\circ \chi^{-1}:\C^n\rightarrow\C^n$ be a holomorphic 
patching matrix   for a vector bundle $E$ over $T\CP^1$
defined on $U\cap\tilde{U}$. 
Restrict $F$ to a section (\ref{minit1}) where the bundle is trivial, and
therefore $F$ can be split (compare (\ref{triv_rel}) in the Appendix)
\be
\label{RH_ward}
F=\t{H}H^{-1},
\ee 
where the matrices $H$ and $\t{H}$ are defined on $M\times \CP^1$ and
are holomorphic in $\pi^{A}$ 
around $\pi^A=o^A=(1, 0)$ and $\pi^A=\iota^A=(0, 1)$ respectively.
As a consequence of $\delta_A F=0$
the splitting  matrices satisfy
\be
\label{definition_of_phiab}
H^{-1}\delta_A H=\widetilde{H}^{-1}\delta_A\widetilde{H}=\pi^{B}\Phi_{AB},
\ee
for some $\Phi_{AB}(x^\mu)$ which does not depend on $\l$. This is because
the RHS and LHS are homogeneous of degree one in $\pi^A$  and holomorphic
around $\lambda=0$ and $\lambda=\infty$ respectively. 
({\bf Exercise}: Prove it starting from the Liouville theorem which says
that any function holomorphic on $\CP^1$ must be constant).
Decomposing
\[
\Phi_{AB}=\Phi_{(AB)}+\varepsilon_{AB}\phi
\]
gives a one--form $A=\Phi_{AB} dx^{AB}$ and a scalar field
$\phi=(1/2)\varepsilon^{AB}\Phi_{AB}$ on the 
complexified Minkowski space, i. e.
\[
{\Phi}_{AB}=
\left (
\begin{array}{cc}
A_u &A_x+\phi\\
A_x-\phi& A_v
\end{array}
\right ).
\]
The Lax pair (\ref{laxpairW}) 
becomes
\[
L_A=\delta_A +H^{-1}\delta_ A H
\]
where  $\delta_A=\pi^B\p_{AB}$,
so that
\[
L_A(H^{-1})=-H^{-1}(\delta_A H) H^{-1}+H^{-1}(\delta_A H) H^{-1}=0
\]
and $\Psi=H^{-1}$ is a solution to the Lax equations regular around
$\lambda=0$.
Let us show explicitly that (\ref{Wardeq0}) holds.
Differentiating both sides to  (\ref{definition_of_phiab}) yields
\[
\delta^A (H^{-1}\delta_A H)=-(H^{-1}\delta^A H )(H^{-1}\delta_A H)
\]
which holds for all $\pi^A$ if
\be
\label{spinor_ward}
D_{A(C}{\Phi^A}_{B)}=0
\ee
where $D_{AC}=\p_{AC}+\Phi_{AC}$. This is the spinor form of
the Yang--Mills--Higgs system (\ref{Wardeq0}).
\koniec
\begin{itemize}
\item To single out the Euclidean reality conditions leading to
non--abelian monopoles (\ref{monopole_R3}) on $\R^3$
with the gauge group $SU(n)$ the vector bundle $E$
must be compatible with the involution (\ref{eucl_invo}). This comes
down to $\mbox{det} F=1$ and
\[
F^*(Z)=F(\tau(Z))
\]
where $Z\in \T$ and $*$ denotes the Hermitian conjugation.
\item 
To single out the Lorentzian reality conditions the bundle must be invariant
under the involution (\ref{lor_invo}). Below we shall demonstrate
how the gauge choices leading to the integrable chiral model (\ref{Wardeq})
can be made at the twistor level.

Let
\[
h:=H(x^{\mu}, \pi^A=o^A), \qquad 
\tilde{h}:=\widetilde{H}(x^{\mu}, \pi^A=\iota^A)
\]
so that
\[
\Phi_{A0}=h^{-1}\p_{A0} h, \qquad \Phi_{A1}=\tilde{h}^{-1}\p_{A1}
\tilde{h}.
\]
The splitting matrices are defined up to a multiple by 
an inverse of 
a non--singular
matrix $g=g (x^{\mu})$ independent on $\pi^A$ 
\[
H\rightarrow H g^{-1}, \qquad \widetilde{H}\rightarrow
\widetilde{H} g^{-1}.
\]
({\bf Exercise.}: Show that
this corresponds to the gauge transformation (\ref{gauge_trans_Ward_m})
of $\Phi_{AB}$).

We choose $g$ such that $\tilde{h}={\bf 1}$ so 
\[
\Phi_{A1}=\iota^A\Phi_{AB}=0
\]
and
\[
\Phi_{AB}=-\iota_B o^C h^{-1}\p_{AC} h,
\]
i. e.
\[
A_x+\phi=A_v=0.
\]
This is the Ward gauge  with $J(x^{\mu})=h$.
In this  gauge  the system (\ref{spinor_ward}) reduces to
\[
{\p^A}_1\Phi_{A0}=0
\]
which is (\ref{Wardeq}). The solution is given by
\[
J(x^{\mu})=\Psi^{-1}(x^{\mu},
\lambda=0)
\] 
where $\Psi=H^{-1}$ is a solution to the  Lax pair.
\item
In the  abelian case  $n=1$ the patching matrix becomes a
function defined on the intersection of two open sets and we can
set  $F=\exp{(f)}$ for some $f$. The non--linear splitting (\ref{triv_rel}) 
reduces to the additive splitting of $f$ which can be carried out explicitly
using the Cauchy integral formula.
The Higgs field is now a function that
satisfies the wave equation and is given by 
formula
\[
\phi=\oint_{\Gamma}\frac{\p f}{\p \omega}\rho\cdot d\rho.
\]
where $\Gamma$ is a real contour in a rational curve
$\omega=x^{AB}\pi_A\pi_B$. If the Euclidean reality conditions are chosen
we recover the Whittaker formula (\ref{whitt_form}).
\item {\bf Exercise}: Find the patching matrix for the holomorphic rank 3 budle $E\rightarrow \T$
corresponding to the one--soliton solution to the Tzitzeica equation (\ref{tzitzeica}).
(Note: the solution to this exercise remains unknown to the author).
\end{itemize}
\section{Dispersionless  systems and deformed mini--twistors}
There is a class of integrable systems in 2+1 and three dimensions which do not fit into the framework described in the last
section. They do not arise from 
(\ref{Wardeq0}) and there is no 
finite--dimensional Riemann--Hilbert problem analogous to (\ref{RH_ward})
which leads to their solutions. These dispersionless integrable systems admit Lax representations which do not involve
matrices, like (\ref{laxpairW}), but instead consist of vector fields.
This leads to curved geometries in the following way.
Consider a 
Lax pair
\be
\label{EWlax}
L_0=W-\lambda V+f_0\frac{\p}{\p\lambda},\quad
L_1=V-\lambda \widetilde{W}+f_1\frac{\p}{\p\lambda},
\ee
where $(W, \widetilde{W}, V)$ are
vector fields on a complex three--manifold $M$ (which generalises the complexified Minkowski space),  and
$(f_0, f_1)$ are cubic polynomials in
$\lambda\in \CP^1$. Assume that the distribution spanned by the Lax pair is integrable in the sense of Frobenius i.e.
\[
[L_0, L_1]=\alpha L_0+\beta L_1
\]
for some $\alpha, \beta$. The twistor space $\T$ is defined to be the quotient
of the total space of the projective spin bundle $P(\spp)\rightarrow M$
by this distribution, i. e.
\[
{\T}=M\times\CP^1/(L_0, L_1).
\]
This is a deformation of $T\CP^1$ (or its region as in general 
the construction is local
in $M$ so $\T$ is taken to be a tubular neighbourhood of a rational curve
corresponding to $p\in M$) which arises if $L_0, L_1$ are given by
(\ref{mini_dist}).

The twistor space is a complex surface containing a three--parameter
family of rational curves $\CP^1$ with self intersection number $2$.
In general $\T$ does not fiber holomorphically over $\CP^1$ which
is a consequence of the presence of $\p/\p \lambda $ terms
in the Lax pair (\ref{EWlax}).

Conversely, given such complex manifold $\T$ one defines $M$ to be
the moduli space of rational curves in 
$\T$ (Kodaira theorems \cite{Kodaira} guarantee that $M$ exists and is 
three complex dimensional).
One can show \cite{Hi82} that $M$
comes equipped with the geometric structure consisting of a conformal 
structure $[h]$, and a compatible torsion--free connection $\nabla$.
The details are as follows:
The points of $M$ correspond 
to rational curves with self--intersection two 
in the complex surface ${\T}$ and
points in  ${\T}$ correspond to null surfaces
in ${M}$ . Recall that the normal bundle $N(L)\rightarrow L$ 
to a submanifold $L\subset \T$ is defined by 
\[
N(L)= \cup_{Z\in L} N_{Z}(L), 
\]
where $
N_{Z}=(T_{Z}{\T})/(T_{Z} L)$ is a quotient vector space.
If $L_p\subset\T$ is the curve corresponding to $p\in M$
then the elements of $T_pM$ correspond to sections of the normal bundle
$N(L_p)$
and as a holomorphic line bundle $N(L_p)\cong \O(2)$ (see Appendix).
 The conformal structure on $M$ arises as
we define  the null vectors at $p$ in ${M}$ to be the
sections of the normal bundle $N(L_p)$ which vanish at some point
to the second order. A section of $\O(2)$ has a form
$V^{AB}\pi_{A}\pi_{B}$ (see Appendix), thus the vanishing condition
$(V^{01})^2- V^{00} V^{11}$ is quadratic and defines $[h]$. 
If $p_1, p_2$ are two points in $M$ which are not null separated,
then the corresponding curves in $\T$ intersect at two points.
If $p_1$ and $p_2$ are infinitesimally close, and thus are joined by
a vector starting from $p_1$, then the corresponding section
of $N(L_1)$ will vanish at two points.
\begin{center}
\includegraphics[width=11cm,height=6cm,angle=0]{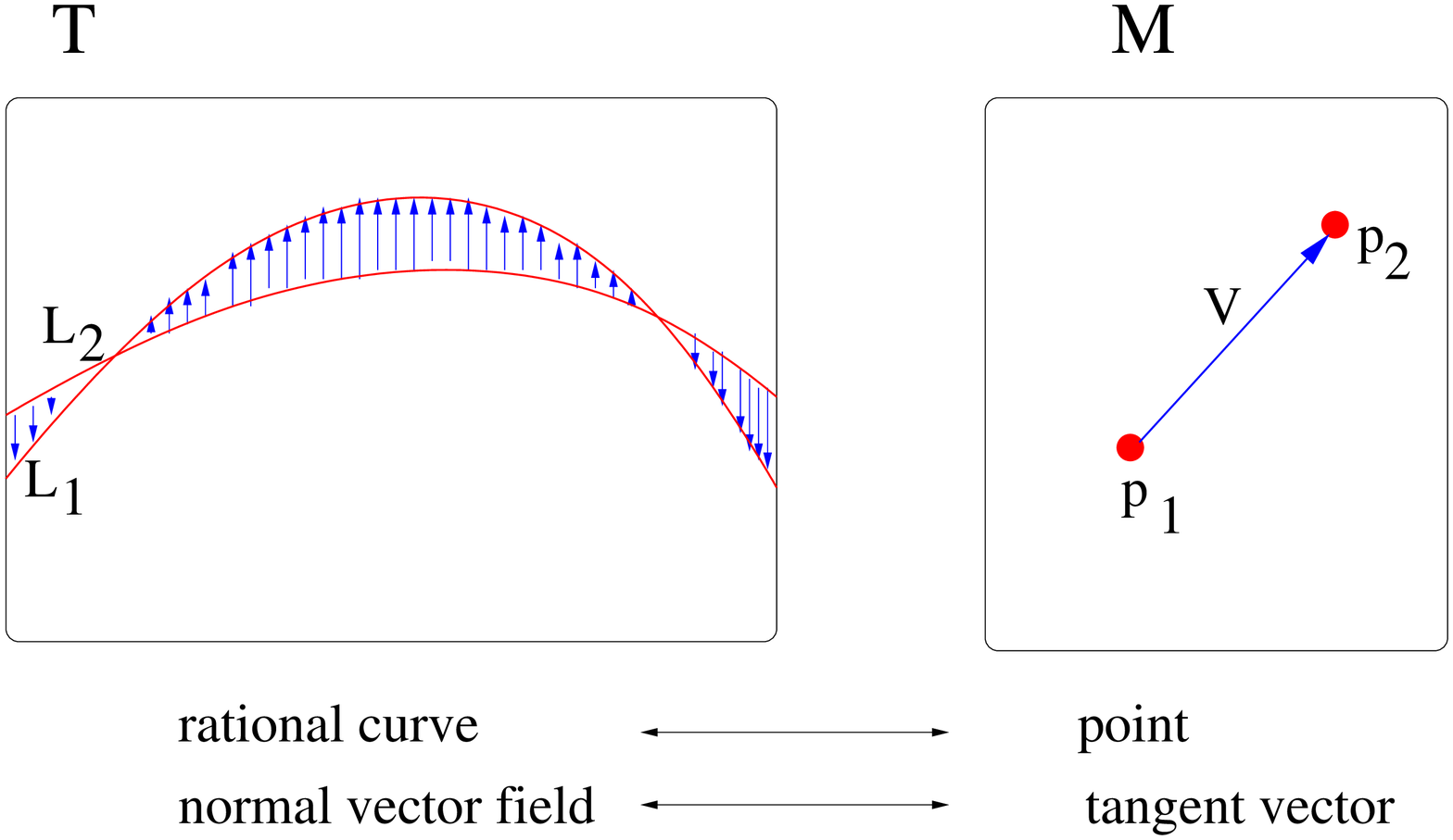}
\end{center}
To define the connection
$\nabla$ we define a direction at $p\in {M}$ to be a one--dimensional
space of sections 
of $\O(2)$  which vanish at two points
$Z_1$ and $Z_2$ in $L_p$. The one--dimensional family of $\O(2)$ curves
in ${\T}$ passing through $Z_1$ and $Z_2$ gives a geodesic
curve in ${M}$ in a given direction and defines $\nabla$. 
In the limiting case
$Z_1=Z_2$ these geodesics are null with respect to $[h]$. This 
compatibility means
that for any choice of $h\in [h]$
\[\nabla h=\omega\otimes h,
\] 
for some one--form $\omega$ on $M$. 
This condition is  
invariant under the conformal rescalings of $h$ if 
\[
h\longrightarrow c^2 \,h, \qquad \omega\longrightarrow \omega
+2 d\;(\ln{(c)}),
\]
where $c$ is a non--zero function on $M$.
Therefore the null geodesics for $[h]$ are also geodesic of $\nabla$ and
thus the pair $([h], \nabla)$ gives a Weyl 
structure on $M$. The Weyl structures
coming from a twistor space satisfy a set of equations generalising 
Einstein equations. This is because
the special surfaces in $M$ corresponding
to points in $\T$ are
totally geodesic with respect to $\nabla$ (if a geodesic is tangent
to a surface at some point then it lies in that surface).
The integrability conditions for the existence of totally
geodesic surfaces is equivalent to 
the conformally invariant Einstein--Weyl equations 
\[
R_{(jk)}=\Lambda h_{(jk)}
\]
where $R_{(jk)}$ is the symmetrised Ricci tensor of the connection $\nabla$,
and $\Lambda$ is some function on $M$.

The Einstein--Weyl equations  admit a Lax formulation
with the Lax pair given by (\ref{EWlax}): If the distribution spanned by 
(\ref{EWlax}) is integrable then there exists a one--form $\omega$ such that 
the metric $h$ given by 
\be
\label{conf_str}
h=V\otimes V-2(W\otimes\widetilde{W}+\widetilde{W}\otimes W)
\ee
and  $\omega$ satisfy
the Einstein--Weyl equations. Any Einstein--Weyl structure 
arises from such a Lax pair \cite{DMT00}.

An example of a dispersionless system which fits into this construction is the 
interpolating integrable system \cite{D08}
\be
\label{uw_eq}
u_y+w_x=0, \quad u_t+w_y-c(uw_x-wu_x)+buu_x=0,
\ee
where $u=u(x, y, t), w=w(x, y, t)$ and $(b, c)$ are constants.
It admits a Lax pair 
\[
L_0=\frac{\p}{\p t}+(cw+bu-\l cu-\l^2)\frac{\p}{\p x}+
b(w_x-\l u_x)\frac{\p}{\p \l}, \quad
L_1=\frac{\p}{\p y}-(cu +\l)\frac{\p}{\p x}-bu_x\frac{\p}{\p\l}
\]
A linear combination of $L_0, L_1$ is of the form (\ref{EWlax}).
The Einstein--Weyl structure associated to (\ref{uw_eq}) is
\begin{eqnarray*}
h&=&(d y-cu\, d t)^2-4(d x-(cw+bu)\,d t)\, d t,\\
\omega&=&-cu_x\,d y+(4bu_x+c^2uu_x -2cu_y)\, d t.
\end{eqnarray*}
({\bf Exercise}: Verify that (\ref{uw_eq}) arises as  $[L_0, L_1]=0$ 
from the given Lax pair.
Use (\ref{conf_str}) to construct the given metric  $h$
from $(u, w)$).
Setting $c=0, b=1$  gives the  
dispersionless Kadomtsev--Petviashvili  equation.
On the twistor level this limit is characterised \cite{DMT00} by the existence of a
preferred section of $\kappa^{-1/4}$ where $\kappa$ is the canonical
bundle of holomorphic two--forms on $\T$.
Another interesting limit
is  ($b=0, c=-1$), where the corresponding 
twistor space fibers holomorphically over $\CP^1$.

There are several approaches to dispersionless integrable systems
in 2+1 dimensions: the Krichever algebro--geometric approach, 
the hydrodynamic reductions developed by Ferapontov
and his collaborators, The Cauchy problem of Manakov--Santini
and the  $\overline{\p}$--formulation of 
Konopelchenko and Martinez Alonso to name a few 
(see \cite{Kri94, Kon3, Bog_kon, FK2,  Manakov_santini}).
The Einstein--Weyl geometry and the associated deformed mini--twistor
theory provide another framework which is coordinate independent, 
and geometric as the solutions are parametrised by complex manifolds with embedded
rational curves.

\section{Summary and outlook}
Twistor theory arose as a non--local attempt to unify
general relativity and quantum mechanics. In this theory
a space time point is a derived object corresponding to rational
curve in some complex manifold. The mathematics behind twistor
theory has its roots in 19th century projective geometry of 
Pl\"ucker and Klein, but it can also be traced back to integral
geometry of Radon and John developed in the first half of the 20th
century. While the twistor programme is yet to have its big impact
on physics (however see \cite{Wi03}), it has lead to methods of solving linear
and non--linear differential equations. In the linear case
one gets nice geometrical interpretations of integral
formulae of Whittaker and John. The twistor
methods of solving nonlinear integrable PDEs are genuinely
new and lead to parametrising `all' solutions by unconstrained holomorphic data. In the case
of the Ward model and  its reductions (as well as 
the anti--self--dual Yang--Mills
equations  \cite{Wa77}
not discussed in this review) the solutions correspond
to holomorphic vector bundles trivial on twistor lines. 
The solutions of dispersionless integrable models (as well
as anti--self--dual conformal equations 
\cite{Pe76} and heavenly equations)
correspond to holomorphic deformations of the complex
structure underlying the twistor space.
  
It is unlikely that all integrable equations fit into one of the
(rather rigid) frameworks (\ref{laxpairW}) or (\ref{EWlax})  presented in this review. It should however
be possible to extend these frameworks, while
keeping their essential features, to incorporate those integrable 
systems which so far have resisted the twistor approach. 

\section*{Appendix}
\setcounter{equation}{0}
\appendix
\def\theequation{\thesection{A}\arabic{equation}}
\subsubsection*{Riemann sphere} 
Two--dimensional sphere $S^2\subset \R^3$ is 
a one-dimensional complex manifold with local coordinates defined by 
stereographic projection. Let $(u_1, u_2, u_3)\in S^2$. 
Define two open subsets covering $S^2$
\[
U=S^2-\{(0, 0, 1)\}, \qquad \t U=S^2-\{(0, 0, -1)\}
\]
and introduce complex coordinates 
$\l$ and $\lt$ on $U$ and $\t U$ respectively 
by
\[
\l=\frac{u_1+iu_2}{1-u_3}, \qquad \lt=\frac{u_1-iu_2}{1+u_3}.
\]
The domain of $\l$ is the whole sphere less the North pole; 
the domain of $\tilde{\l}$  is the whole sphere less the South pole. 
On the overlap $U_0\cap U_1$ we have 
$\lt=1/\l$ which is a holomorphic function.
The resulting complex manifold is called $\CP^1$. It also arises
as the quotient of $\C^{2}$ by the equivalence relation 
$$
(\pi_0, \pi_1)\sim 
(c \pi_0, c\pi_1)\qquad \mbox{for some}\;\;c\in 
\C^*\, .
$$ 
The homogeneous coordinates $\pi_A$ label the points uniquely, up 
to an overall non--zero complex scaling factor.  In this approach  
complex manifold
structure on  $\CP^1$ is introduced by using the  
{inhomogeneous coordinates}.  
On the open set $U$ in which $\pi_1\neq 0$, we define $\lambda=\pi_0/\pi_1$
and on the open set $\tilde{U}$ with  $\pi_0\neq 0$
we set $\lt=\pi_1/\pi_0$ so that $\lt=1/\l$ on the overlap.
\subsubsection*{Holomorphic vector bundles}
A holomorphic vector bundle of rank $n$  over a complex manifold $\T$ is
a complex manifold $E$, and a holomorphic projection 
$\pi:E\rightarrow \T$ such that
\begin{itemize}
\item For each $z\in \T$, $\pi^{-1}({z})$ is an $n$-dimensional 
complex vector space.
\item 
Each point $z\in \T$ has a neighbourhood $U_\a$ and a homeomorphism
$\chi_\a$ such that the diagram
\[\begin{array}{rcccc}
   & & \chi_\a & & \\
  {\pi}^{-1}(U_\a) & & \cong & & U_\a\times{\C^n} \\
  \pi & \searrow & & \swarrow & \\
   & & U_\a & & 
  \end{array}
\]
is commutative.
\item The patching matrix
$F_{\a\beta}:=\chi_\beta\circ {\chi_\a}^{-1}:U_\a\cap U_\beta
\rightarrow {\rm GL}(n,\C)
$ 
is a holomorphic map to the space of invertible $n\times n$ matrices.
\end{itemize}

The product $E=\T\times\C^n$ is called a trivial vector bundle.
The bundle is trivial,  iff there exist holomorphic 
splitting matrices $H_\a:U_\a\to {\rm GL}(n,\C)$ such that 
\be
\label{triv_rel}
F_{\a\beta}=H_{\beta} H_{\a}^{-1}\, .
\ee

We shall give examples of holomorphic line bundles (i.e. vector bundles
with $n=1$) over
$\CP^1$. First define a tautological line bundle 
\[
\O(-1)=\{(\l, (\pi_0, \pi_1))\in \CP^1\times\C^2|\l=\pi_0/\pi_1\}.
\]
Representing the Riemann sphere as 
the projective line gives the projection $\C^2\to \CP^1$.  
The fibre above the point with coordinate $[\pi]$ is the 
one-dimensional line  $c\pi$ through the origin  in $\C^2$ 
containing the the point $(\pi_0, \pi_1)$.
The transition function for this bundle is $F=\l$. ({\bf Exercise}: Show it).
Other line bundles can be obtained  by algebraic operations:
\[
\O(-m)=\O(-1)^{\otimes m}, \qquad \O(m)=\O(-m)^*, \qquad\O=\O(-1)\otimes\O(1), 
\qquad m\in \NN.
\]
The transition function for 
${\cal O}(m)$ is $F=\l^{-m}$ on $U\cap \tilde{U} \cong {\C}^*$.

The line bundles ${\cal O}(m)$  for any $m\in \Z$ are building blocks for
all other vector bundles over the Riemann sphere. This is a consequence of
the Birkhoff--Grothendieck theorem which states that a rank $n$
holomorphic vector bundle $E\rightarrow \CP^1$ 
is isomorphic to a direct sum of line bundles ${\cal O}(m_1)\oplus \cdots \oplus {\cal 
O}(m_n)$ for some integers $m_i$.
\subsubsection*{Holomorphic sections}
A holomorphic section
of a vector bundle $E$ over $\T$ is a holomorphic map
$s:\T\rightarrow E$ such that $\pi\circ s={\rm id}_\T$.
The local description is given by a collection of holomorphic maps
$s_\a:U_\a\rightarrow\C^n$
\[
z\longrightarrow (z, s_\a(z)), \qquad \mbox{for}\;\;z\in U_\a.
\]
with the transition
rule   $s_\beta(z)=F_{\a\beta}(z)s_\a(z)$.

A global holomorphic section of 
the line bundle  $\O(m)$ is given by 
functions $s$ and $\t s$ on ${\C}$ 
holomorphic in $\lambda$ and $\tilde{\lambda}$ respectively and 
related by
$$s (\l)=\l^m \t s (\tilde{\l})$$
on the overlap ${\C}^{\ast}$. Expanding these functions as power series in
their respective local coordinates, and using the fact that 
$\tilde{\l}=\l^{-1}$ 
 and hence the space of holomorphic sections of ${\cal O}(m)$ 
is $\C^{m+1}$ if $m>0$. There are no global holomorphic sections if $m<0$.
A global holomorphic section of $\O(m), m\geq 0$ 
is the same as a global function on $\C^2$
homogeneous of degree $m$ (a polynomial). If $m>0$ such function is of the 
form
\[
f([\pi])= V^{AB\ldots C} \pi_A\pi_B\cdots\pi_C
\] 
for some symmetric object $V^{AB\ldots C}$. 

Holomorphic vector fields on $\CP^1$ are sections of the holomorphic 
tangent bundle $T\CP^1$.
Using 
\[
\frac{\p}{\p\l}=-\l^{-2}\frac{\p}{\p\lt},
\]
and absorbing the minus signs into the local trivialisations, we deduce that
$
T\CP^1=\O(2).
$
({\bf Exercise}: Consider a general section of $\O(2)\rightarrow \CP^1$
given by the local form (\ref{twistor_curve}) 
where $(v, x, u)$ and $(\eta, \lambda)$ are
complex. Show that this section is invariant under (\ref{eucl_invo}) if
$x\in\R$ and $u=-\ov{v}$. Thus deduce (\ref{hol_sec_point})).

\end{document}